\documentclass[aps,prd,twocolumn,floatfix,preprintnumbers,showpacs]{revtex4}
\usepackage{amssymb,amsmath,bm,graphicx,longtable}
\usepackage{mathrsfs,enumerate}

\begin{document}

\preprint{CHIBA-EP-183/KEK Preprint 2010-8}

\title{Jackiw-Nohl-Rebbi two-instanton as a source of  magnetic monopole loop
}

\author{Nobuyuki Fukui$^{1}$}
\email{n.fukui@graduate.chiba-u.jp}

\author{Kei-Ichi Kondo$^{1}$}
\email{kondok@faculty.chiba-u.jp}

\author{Akihiro Shibata$^{2}$}
\email{akihiro.shibata@kek.jp}

\author{Toru Shinohara$^{1}$}
\email{sinohara@graduate.chiba-u.jp}

\affiliation{$^1$Department of Physics,  Graduate School of Science, Chiba University, Chiba 263-8522, Japan
\\
$^2$Computing Research Center, High Energy Accelerator Research Organization (KEK) 
and Graduate Univ. for Advanced Studies (Sokendai), Tsukuba  305-0801, Japan
}

\date{\today}

\begin{abstract}
We demonstrate that a Jackiw-Nohl-Rebbi solution as the most general two-instanton generates a circular loop of  magnetic monopole in four-dimensional Euclidean SU(2) Yang-Mills theory, in contrast to the one-instanton solution in the regular gauge for which no such magnetic monopole loops exist. 
These results together with our previous result indicate that two-instanton solution and two-meron solution with the same asymptotic behavior in the long distance are responsible for quark confinement based on the dual superconductivity picture. 
\end{abstract}

\pacs{12.38.Aw, 21.65.Qr}

\maketitle

\section{Introduction}\label{sec:intro}

The quark confinement criterion {\it a la} Wilson \cite{Wilson74} is given by the area decay of the Wilson loop average $\langle W_C[\mathscr{A}] \rangle$, i.e., the vacuum expectation value of the Wilson loop operator $W_C[\mathscr{A}]$ written in terms of the Yang-Mills field \cite{YM54} $\mathscr{A}_\mu(x)$. 
In order to clarify the mechanism of quark confinement, it would be efficient to construct ensembles of gauge field which exhibit confinement.  
One anticipates that a class of solutions of the classical Yang-Mills equation may give the dominant contribution to quark confinement.   
Among them, especially, topological (soliton) solutions \cite{MS} with non-trivial topological charge $Q_P$, i.e., the Pontryagin index,  are good candidate for a first examination. 

The well-known topological solutions are instantons \cite{BPST75,CF77,tHooft76,Wilczek77,Witten79,JNR77,ADHM78} and merons \cite{AFF76,AFF77,Actor79}:
The instanton is a solution of the self-duality equation ${}^*F= \pm F$ (first order differential equation) and has an integer-valued topological charge $Q_P= \pm1, \pm 2, \cdots$, while the meron is a solution of the original second order differential equation obtained without imposing the self-duality condition and has a half-integer- or integer-valued topological charge $Q_P=\pm 1/2, \pm 1, \cdots$.
Remarkably, the instanton has a finite Euclidean action proportional to the topological charge $S=8\pi^2 |Q_P|/g^2$ due to self-duality, while the meron has the logarithmically divergent Euclidean action due to short distance singularity, since the meron has no collective coordinate corresponding to the size. 
Therefore, the requirement of finite action inevitably excludes the meron solution from the candidates.  However, it has been pointed out in \cite{CDG78} that merons can be responsible for the occurrence of the confinement phase, if we take into account the free energy based on the action-entropy argument, since the logarithmic divergent action is comparable with the entropy associated with the meron configurations calculable from the integration measure in the path-integral formulation.

It is believed that a promising mechanism for   quark confinement is a dual superconductivity \cite{dualsuper,Polyakov77}. 
For this mechanism to work, however, magnetic monopoles must exist and be  condensed to cause the dual superconductivity. 
In this picture, magnetic monopoles are considered to be the most important degrees of freedom relevant to confinement in the dual description. 
Therefore, we are lead to estimate how magnetic monopoles contribute  to the Wilson loop average to fulfill the confinement criterion \cite{Kondo08b}.

On the other hands, we can ask which configuration of the Yang-Mills field can be the source for such magnetic monopoles relevant to  confinement in the dual description. 
The simplest configuration examined first was the one instanton solution \cite{BPST75}.
However, it has been confirmed in  \cite{CG95,BOT97,BHVW01,KFSS08} that magnetic monopole loops are not generated from one-instanton configuration, as briefly reviewed in the Introduction of our previous paper \cite{KFSS08}. 
Note that the magnetic monopole expressed by the current $k$ is a topological object of co-dimension 3, therefore, it is a one-dimensional object (a closed current due to the topological conservation $\delta k=0$) in the four dimensional space, while it is a point-like object in the three dimensional space. 

In previous papers, we have shown that two meron solution with a unit total topological charge $|Q_P|=1$ leads to circular loops of magnetic monopole  joining a pair of merons in an analytical way \cite{KFSS08} and a numerical way \cite{SKKISF09},
although one-instanton and two-merons have the same total topological charge. 
This result is in good agreement with the numerical result \cite{MN02} obtained in the Laplacian Abelian gauge. 
 
In this paper we examine the two-instanton solution of Jackiw-Nohl-Rebbi (JNR) \cite{JNR77} from the viewpoint raised above. 
In the conventional studies on  quark confinement, the multi-instanton solution of 't Hooft type \cite{tHooft76} has been used extensively to see the interplay between instantons and magnetic monopoles \cite{STSM96,BOT97,RT01,BH03}. 
However, the 't Hooft instanton is not the most general instanton solutions except for the one-instanton case in which the 't Hooft one-instanton agrees with the well-known one-instanton solution in the singular gauge. 
In contrast, the JNR two-instanton solution is the most general two-instanton solution with the full collective coordinates (moduli parameters), while the 't Hooft two-instanton solution \cite{tHooft76} is obtained as a special limit of the JNR solution. 
We demonstrate in a numerical way that a circular loop of magnetic current $k$ is generated for a JNR two-instanton solution.
In addition, we present the configuration of the color field which plays the crucial role in our formulation.  
The implications of this result for quark confinement will be discussed in the final section. 

Incidentally, the JNR two instanton was used to study the relationship between dyonic instantons as a supertube connecting two parallel D4-branes and the magnetic monopole string loop as the supertube cross-section  in (4+1) dimensional Yang-Mills-Higgs theory \cite{KL03}, since dyonic instantons of 't Hooft type do not show magnetic string  and D4-branes meet on isolated points, instead of some loop.  
These facts became one of the motivations to study the JNR solution from our point of view.

\section{Magnetic monopole current on a lattice}

It is a nontrivial question how to realize the magnetic monopole in the pure Yang-Mills theory without matter field, while in the Yang-Mills-Higgs system such as the Georgi-Glashow model, the magnetic monopole in a non-Abelian gauge theory has been constructed long ago \cite{tHP74}.

In pure Yang-Mills theory in the absence of matter fields, two methods are currently known (to the best of our knowledge) for extracting magnetic monopole degrees of freedom: 
\begin{enumerate}

\item
  Abelian projection 
due to 't Hooft \cite{tHooft81}  

\item
  Field decomposition 
due to Cho, Duan and Ge,   Faddeev and Niemi, Shabanov   \cite{DG79,Cho80,FN99,Shabanov99}

\end{enumerate}
 The first method (Abelian projection) has succeeded to exhibit the Abelian dominance \cite{SY90,AS99} and magnetic monopole dominance \cite{SNW94} for quark confinement by adopting the Maximally Abelian (MA) gauge \cite{KLSW87}.
See e.g., \cite{CP97} for a review.
However, the following questions were raised for results obtained in MA gauge, which are fundamental questions to be answered to establish the gauge-independent dual superconductivity in Yang-Mills theory. 
(1)
How to extract the {``Abelian'' part} responsible for quark confinement  from the non-Abelian gauge theory in the {gauge-independent way}.
(2) 
How to define the {magnetic monopole} to be condensed in  Yang-Mills theory in the {gauge-invariant way} even in absence of any  fundamental scalar field, in sharp contrast to the Georgi-Glashow model.

The MA gauge is reproduced as a special limit of the second method. In fact, the first method is nothing but a gauge-fixed version of the second method.  
In other words, the second method is a manifestly gauge-covariant reformulation of the first one.
Therefore, the second method enables us to answer the above  questions.  
From this viewpoint, the second method has been developed in a series of our papers \cite{KMS06,KMS05,Kondo06,KKMSSI05,IKKMSS06,SKKMSI07,SKKMSI07b,{KKSSI09}}, for $SU(2)$ gauge group and \cite{KSM08,KSSMKI08,SKS09} for $SU(3)$ or $SU(N)$ gauge group. 
In the second method, a magnetic monopole can be defined in a manifestly gauge-invariant way using new variables obtained from the original Yang-Mills field by change of variables.
Moreover, the Wilson loop operator can be exactly rewritten in terms of the magnetic monopole defined in this way through a non-Abelian Stokes theorem \cite{Cho00,Kondo08,KS08}.

In order to define the magnetic monopole on a lattice, we recall the second method based on a non-linear change of variables in a continuum $SU(2)$ Yang-Mills theory.
We introduce a color field ${\bf n}(x)$ with a unit length:
\begin{gather}
 {\bf n}(x)=n^A(x)T^A,\quad\left(T^A :=\sigma_A/2\right)
\\
 n^A(x)n^A(x)=1 ,
\end{gather}
where $\sigma_A\ (A=1,2,3)$ are Pauli matrices. 

The color field is determined by imposing a condition which we call the reduction condition.
A reduction condition is given by minimizing the functional 
\begin{equation}
F_{\text{red}} 
:= \int d^4x \frac12 \text{tr} [\{D_\mu[{\bf A}]{\bf n}(x)\}^2   ] .
\end{equation}
The local minima are given by the differential equation  which we call the reduction differential equation (RDE) \cite{KFSS08}:
\begin{equation}
 - D_\mu[{\bf A}]D_\mu[{\bf A}]{\bf n}(x)=\lambda(x){\bf n}(x) .
\label{RDE}
\end{equation}

In this theory, a composite field
\begin{equation}
 {\bf V}_\mu(x) :=
c_\mu(x) {\bf n}(x)-ig^{-1}\left[\partial_\mu{\bf n}(x), {\bf n}(x)\right]
\label{V}
\end{equation}
play an important role 
where $c_\mu(x) :=2 \text{tr}\left({\bf n}(x){\bf A}_\mu(x)\right)$.

For instance, the field strength of ${\bf V}_\mu(x)$ is parallel to ${\bf n}(x)$:
\begin{align}
 {\bf F}_{\mu\nu}[{\bf V}]
 &=\partial_\mu{\bf V}_\nu
  -\partial_\nu{\bf V}_\mu
  -ig\left[{\bf V}_\mu, {\bf V}_\nu\right]\notag\\
 &=\left\{\partial_\mu c_\nu
         -\partial_\nu c_\mu
         +2ig^{-1}
          \text{tr}
          \big({\bf n}
                \left[\partial_\mu{\bf n}, \partial_\nu{\bf n}\right]
          \big)
   \right\}{\bf n}\notag
\\
 &:= G_{\mu\nu}{\bf n}.
\end{align}
Thus, we can define the gauge-invariant field strength $G_{\mu\nu}(x) = 2\text{tr}({\bf n}  {\bf F}_{\mu\nu}[{\bf V}]) $ and the gauge invariant monopole current as 
\begin{equation}
 k^\mu(x)
 :=  \partial_\nu\!\,^\ast G^{\mu\nu}(x)\label{k}
 =  \frac{1}{2}\epsilon^{\mu\nu\rho\sigma}\partial_\nu G_{\rho\sigma}(x) .
\end{equation}
Once the reduction condition is solved, thus,  
we can obtain the monopole current $k^\mu$ from the original gauge field ${\bf A}_\mu$.
The gauge-invariant magnetic charge is defined  
\begin{equation}
 q_m 
 :=  \int d^3 \sigma_\mu k^\mu(x)  ,
\end{equation}
in a Lorentz (or Euclidean rotation)  invariant way \cite{Kondo08b}.

In this paper, we  carry out the procedures explained in the above in a numerical way. We use the lattice regularization for numerical calculations where the link variable $U_{x,\mu}$ is related to a gauge field in a continuum theory by
\begin{equation}
 U_{x,\mu}=\text{P}
            \exp
            \left\{ig\int_x^{x+a\hat{\mu}}dy{\bf A}_\mu(y)\right\} ,
 \label{U}
\end{equation}
where P represents a path-ordered product, $a$ is a lattice spacing and
$\hat{\mu}$ represents the unit vector in the $\mu$ direction.
The lattice version of the reduction functional in $SU(2)$ Yang-Mills theory is given by  
\begin{equation}
 F_{\text{red}}[{\bf n},U]
 =\sum_{x,\mu}
  \left\{1-4\,\text{tr}
          \left(U_{x,\mu}{\bf n}_{x+a\hat{\mu}}U_{x,\mu}^\dagger{\bf n}_x\right)
          /\text{tr}\left({\bf 1}\right)
  \right\},
\label{reduction}
\end{equation}
where ${\bf n}_x$ is a unit color field  on a site $x$,
\begin{equation}
 {\bf n}_x=n_x^AT^A,\quad n^A_xn^A_x=1 .
 \label{lattice-n}
\end{equation}

We introduce the Lagrange multiplier $\lambda_x$ to incorporate the constraint of unit length for the color field (\ref{lattice-n}). 
Then the stationary condition for the reduction functional is given by
\begin{equation}
  \frac{\partial }{\partial n_x^A} \left\{ F_{\text{red}}[{\bf n},U] - \frac12  \sum_{x} \lambda_x (n_x^A n_x^A -1) \right\}  =0 .
\end{equation}
When $F_{\text{red}}$ takes a local minimum for a given and fixed configurations $\{U_{x,\mu}\}$, therefore, a Lagrange multiplier $\lambda_x$  satisfies
\begin{equation}
 W_x^A=\lambda_x n_x^A
\label{RDEonLattice1},
\end{equation}
and the color field $n^A_x$ satisfies
\begin{equation}
  n^A_xn^A_x=1 ,
\end{equation}
where
\begin{align}
  &W_x^A=4\sum_{\mu=1}^4\text{tr}
         \Big(U_{x,\mu}{\bf n}_{x+a\hat{\mu}}U_{x,\mu}^\dagger T^A\notag\\
  &\hspace{2cm}
        +U_{x-a\hat{\mu},\mu}T^A
               U_{x-a\hat{\mu},\mu}^\dagger{\bf n}_{x-a\hat{\mu}}
         \Big)/\text{tr}\left({\bf 1}\right)\label{W_x^A}.
\end{align}
Eq.\eqref{RDEonLattice1} is a lattice version of the reduction differential equation (RDE).
We are able to eliminate the Lagrange multiplier to rewrite (\ref{RDEonLattice1}) into
\begin{equation}
 n_x^A=\frac{W^A_x}{\sqrt{W^B_xW^B_x}} .
\label{RDEonLattice2}
\end{equation}
A derivation of this equation is given in Appendix A.
The color field configurations $\{{\bf n}_x\}$  are obtained by solving  \eqref{RDEonLattice2} in a numerical way.

After obtaining the $\{{\bf n}_x\}$ configuration for  given configurations $\{U_{x,\mu}\}$ in this way, we introduce a new link variable $V_{x,\mu}$ on a lattice corresponding to the gauge potential (\ref{V}) by
\begin{align}
 V_{x,\mu}=& \frac{L_{x,\mu}}
                 {\sqrt{\displaystyle{\frac{1}{2}}
                        \ \text{tr}\!
                        \left[L_{x,\mu}L_{x,\mu}^\dagger\right]}} ,
\nonumber\\
 L_{x,\mu}:=& U_{x,\mu}+{\bf n}_xU_{x,\mu}{\bf n}_{x+a\hat{\mu}} .
\end{align}
Finally, the monopole current  $k_{x,\mu}$ on a lattice
is constructed  as 
\begin{equation}
 k_{x,\mu}
 =\sum_{\nu,\rho,\sigma}\frac{\epsilon_{\mu\nu\rho\sigma}}{4\pi}
  \frac{\Theta_{x+a\hat{\nu},\rho\sigma}[{\bf n},V]
       -\Theta_{x,\rho\sigma}[{\bf n},V]}{a} ,
 \label{definition_of_k}
\end{equation}
through the angle variable of the plaquette variable
\begin{align}
 &\Theta_{x,\mu\nu}[{\bf n},V]\notag\\
 &=a^{-2}\arg
  \Big(\text{tr}
        \left\{\left({\bf 1}+{\bf n}_x\right)
               V_{x,\mu}V_{x+a\hat{\mu},\nu}
               V_{x+a\hat{\nu},\mu}^\dagger V_{x,\nu}^\dagger
        \right\}\notag\\
 &\hspace{6.2cm}/\text{tr}\left({\bf 1}\right)
  \Big) .
\end{align}
In this definition, $k_{x,\mu}$ takes an integer value \cite{KKMSSI05,IKKMSS06}.

To obtain the $\{{\bf n}_x\}$ configuration   satisfying \eqref{RDEonLattice2}, we recursively apply \eqref{RDEonLattice2} to ${\bf n}_x$ on each site $x$
and update it keeping ${\bf n}_x$ fixed at a boundary $\partial V$ of a finite lattice $V$
until $F_{\text{red}}$ converges.
Since we calculate the $\{k_{x,\mu}\}$ configuration for the instanton
configuration in this paper, we need to decide a boundary condition of the $\{{\bf n}_x\}$ configuration in the instanton case.
We recall that the instanton configuration
approaches  a pure gauge at infinity:
\begin{equation}
 g{\bf A}_\mu(x)\rightarrow ih^\dagger(x)\partial_\mu h(x) + O(|x|^{-2}) .
\label{BehaviorOfA}
\end{equation}
Then,  ${\bf n}(x)$ as a solution of the reduction condition is supposed to behave asymptotically 
\begin{equation}
 {\bf n}(x)\rightarrow h^\dagger(x)T_3 h(x) + O(|x|^{-\alpha}) ,
\label{BehaviorOfn}
\end{equation}
for a certain value of $\alpha>0$.
Under this idea, we adopt a boundary condition as
\begin{equation}
 {\bf n}_x^\text{bound} :=h^\dagger(x)T_3 h(x) , \ x \in \partial V.
\end{equation}
In practice,
we start with an initial state of the $\{{\bf n}_x\}$ configuration:
${\bf n}_x^\text{init}=h^\dagger(x)T_3 h(x)$ for $x \in V$. Then, we 
repeat updating ${\bf n}_x$ on each site $x$ according to \eqref{RDEonLattice2} except for the configuration ${\bf n}_x^\text{bound}$ on the boundary $\partial V$.

It should be remarked that these asymptotic forms (\ref{BehaviorOfA}) and (\ref{BehaviorOfn}) satisfy the RDE asymptotically in the sense that
\begin{equation}
  D_\mu[{\bf A}]{\bf n}(x) \rightarrow 0  \quad (|x| \rightarrow \infty ),
\end{equation}
together with
\begin{equation}
  \lambda(x) \rightarrow 0 \quad (|x| \rightarrow \infty),
\end{equation}
which is necessary to obtain a finite value for the reduction functional \cite{KFSS08}
\begin{equation}
F_{\text{red}} 
= \int d^4x \frac12  \lambda(x) < \infty  .
\end{equation}

\section{One instanton in the regular gauge}

\begin{figure*}[htbp]
 \unitlength=0.001in
 \begin{picture}(7000,4600)(0,0)
  \put(0,2600){\includegraphics[trim=0 0 0 0, width=90mm]%
              {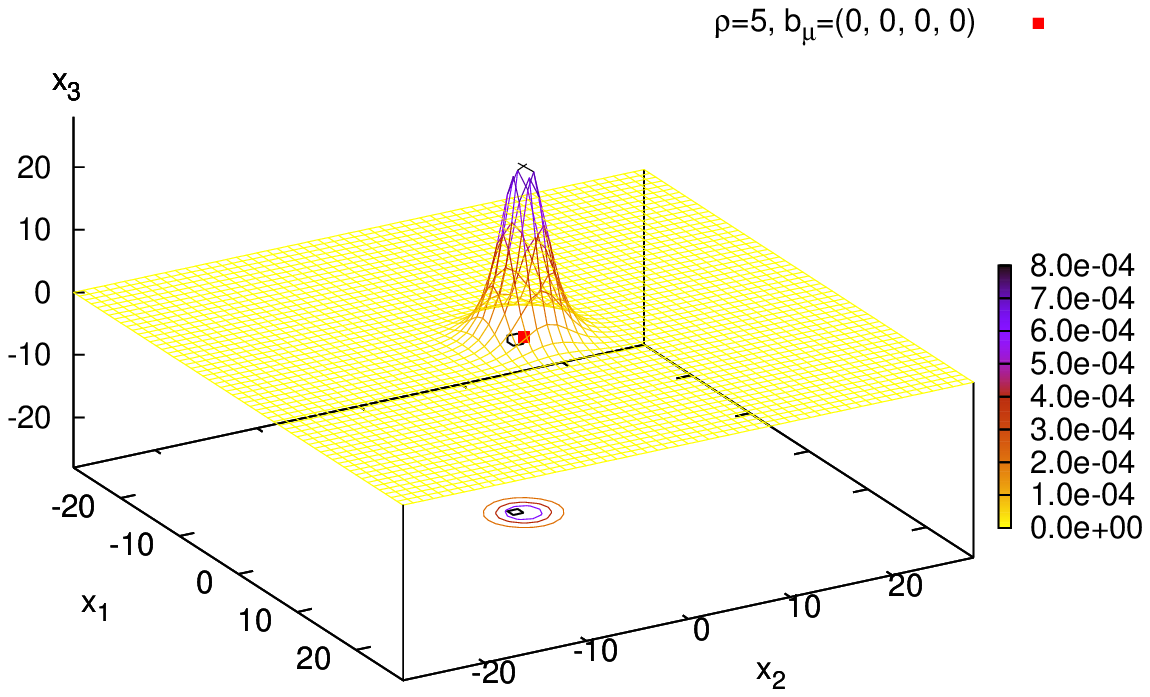}}%
  \put(1600,2600){(a)}
  \put(3500,2600){\includegraphics[trim=0 0 0 0, width=90mm]%
                 {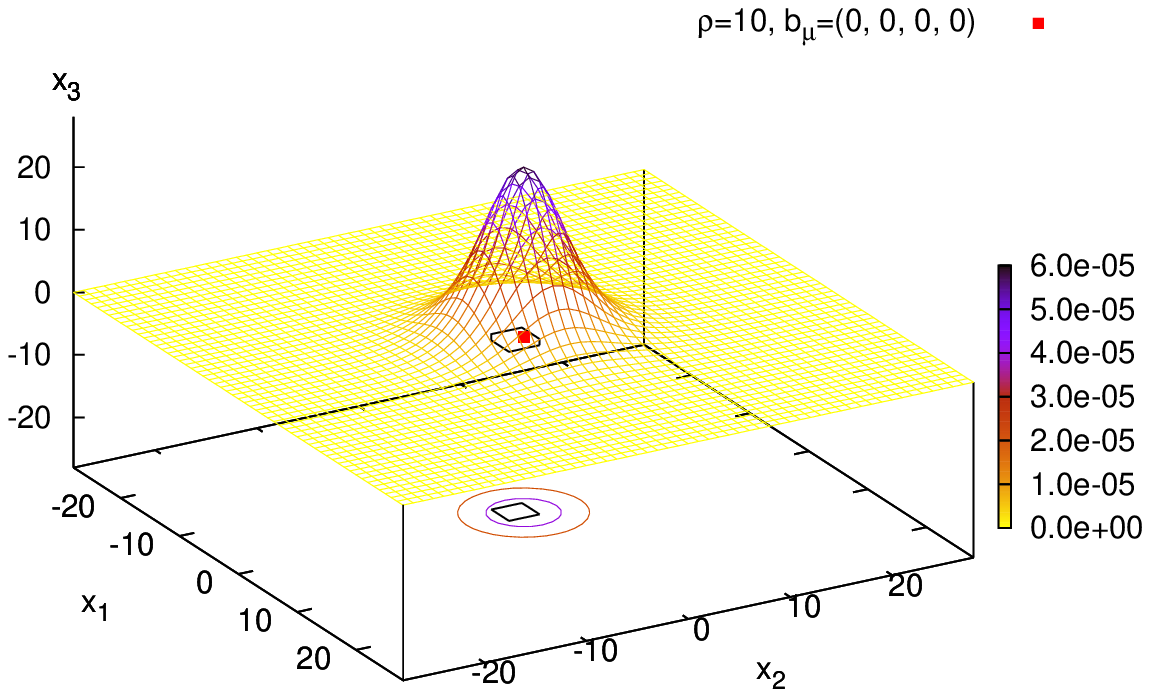}}%
  \put(5100,2600){(b)}
  \put(0,150){\includegraphics[trim=0 0 0 0, width=90mm]%
           {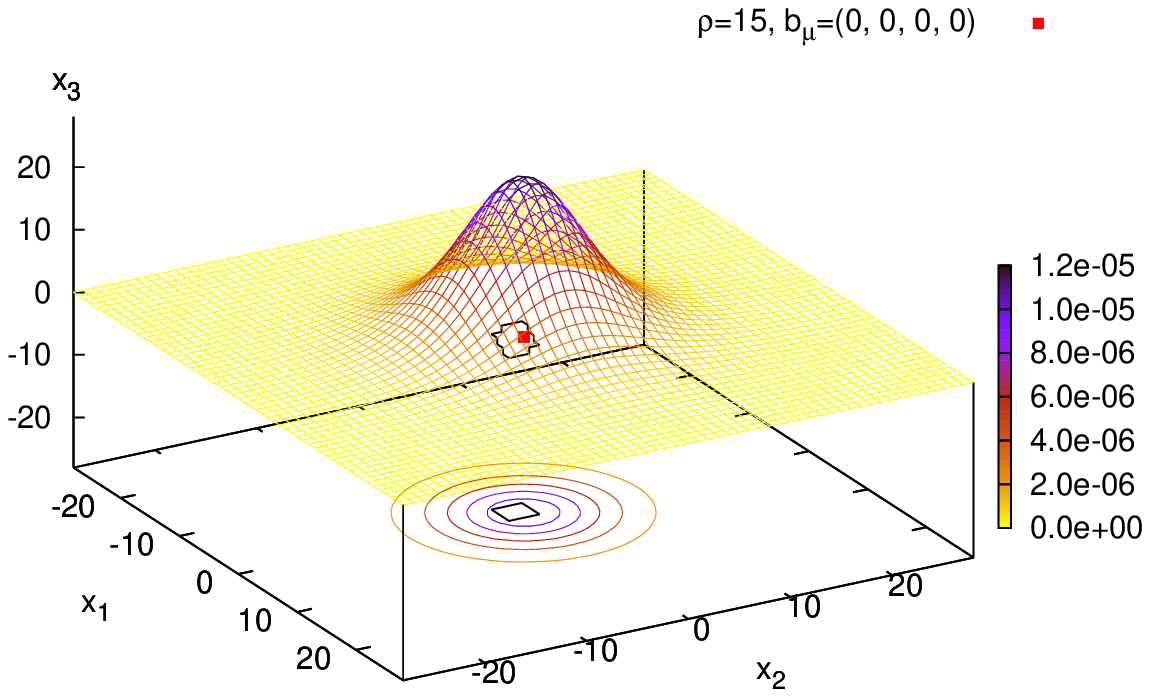}}%
  \put(1600,150){(c)}
  \put(3500,150){\includegraphics[trim=0 0 0 0, width=90mm]%
              {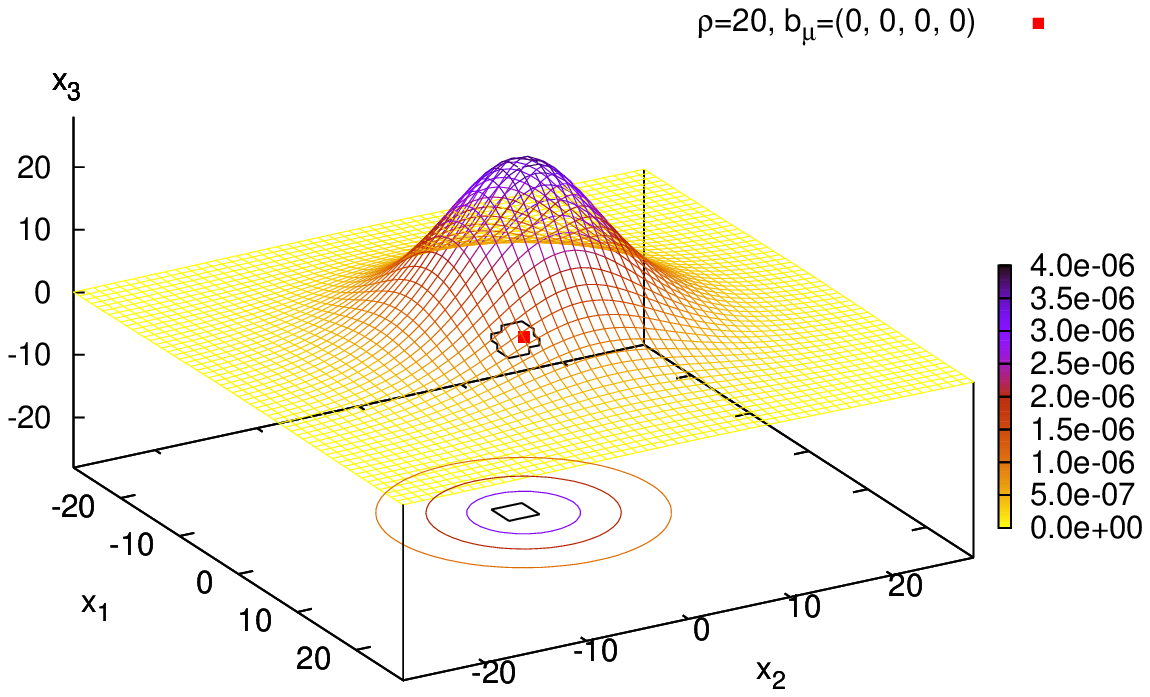}}%
  \put(5100,150){(d)}
 \end{picture}
 \caption{One instanton in the regular gauge and the associated magnetic--monopole current $k_{x,\mu}$ for various choice of size parameter $\rho$: (a) $\rho=5a$, (b) $\rho=10a$,
          (c) $\rho=15a$ and (d) $\rho=20a$.
The grid shows an instanton charge density $D_x$ on $x_1$-$x_2$ ($x_3=x_4=0$) plane.
The black line on the base shows the magnetic monopole loop projected and colored lines shows a contour plot of the instanton charge density.
Figures are drawn in units of $a$.
}
 \label{fig:one_loop}
\end{figure*}

The one-instanton solution in the regular (or nonsingular) gauge is specified by a constant four-vector representing the center $(b^1,b^2,b^3,b^4) \in \mathbb{R}^4$
and a positive real constant representing the size (width) $\rho \ge 0$:
\begin{equation}
 g{\bf A}_\mu(x)= T^A \eta_{\mu\nu}^{A(+)}
                 \frac{2(x^\nu-b^\nu)}{|x-b|^2+\rho^2}, 
\end{equation}
where 
$|x|^2=x_\mu x_\mu$ is the standard Euclidean norm and  $\eta_{\mu\nu}^{A(\pm)}$ is the symbol defined by
\begin{equation}
 \eta_{\mu\nu}^{A(\pm)}
 =\epsilon_{A\mu\nu4}\pm\delta_{A\mu}\delta_{\nu4}\mp\delta_{A\nu}\delta_{\mu4}.
\end{equation}

In this case, we obtain from \eqref{BehaviorOfA} and \eqref{BehaviorOfn}:
\begin{equation}
 h(x)=\frac{x_\mu}{|x|}{e}_\mu,\ 
 \left(e_\mu\equiv(-i\sigma_i,\bm{1})\right) ,
\end{equation}
where $\bm{1}$ is a $2 \times 2$ unit matrix, and
\begin{align}
 &h^\dagger(x)T_3 h(x)=\frac{2\left(x_1x_3-x_2x_4\right)}{x^2}T_1\notag\\
 &\hspace{3cm}+\frac{2\left(x_1x_4+x_2x_3\right)}{x^2}T_2\notag\\
 &\hspace{3cm}+\frac{-x_1^2-x_2^2+x_3^2+x_4^2}{x^2}T_3.
\end{align}
This exactly agrees with the standard Hopf map \cite{Hopf31}.

The topological charge density is maximal at the point $x=b$ and decreases algebraically with the distance from this point in such a way that the instanton charge $Q_V$ inside the finite lattice $V=[-aL,aL]^4$ reproduces  the total instanton charge $Q_P=1$. 
We construct the instanton charge $Q_V$ on a lattice  from the  configuration of link variables $\{U_{x,\mu}\}$ according to 
\begin{align}
 Q_V=& a^4 \sum_{x\in \{ V-\partial V \} }D_x,
\\
 D_x :=& \frac{1}{2^4}\frac{a^{-4}}{32\pi^2}
     \sum_{\mu,\nu,\rho,\sigma=\pm1}^{\pm4}
     \hat{\epsilon}^{\mu\nu\rho\sigma} {\rm tr}(\mathbf{1}-U_{x,\mu\nu}U_{x,\rho\sigma}) ,
\\
 U_{x,\mu\nu}=& U_{x,\mu}U_{x+a\hat{\mu},\nu}
              U_{x+a\hat{\nu},\mu}^\dagger U_{x,\nu}^\dagger ,
\end{align}
where $D_x$ is a lattice version of the instanton charge density, and
$V-\partial V$ represents the volume without a boundary, and $\hat{\epsilon}$ is related to
the usual $\epsilon$ tensor by
\begin{align}
 &\hat{\epsilon}^{\mu\nu\rho\sigma}
 :=\text{sgn}(\mu)\text{sgn}(\nu)\text{sgn}(\rho)\text{sgn}(\sigma)
  \epsilon^{|\mu||\nu||\rho||\sigma|},\notag\\
 &\hspace{4cm} \text{sgn}(\mu):=\frac{\mu}{|\mu|} .
\end{align}

Our interest is the support of $k_{x,\mu}$, namely, a set of links $\{ x,\mu \}$ on which $k_{x,\mu}$ takes non-zero values $k_{x,\mu} \not= 0$.  This expresses the location of the magnetic monopole current generated for a given instanton configuration. 
By definition \eqref{definition_of_k}, the number of configurations $\{ k_{x,\mu} \}$ are $(2L)^4 \times 4$.
(The number of configurations $\{ k_{x,\mu} \}$ is not equal to $(2L+1)^4\times 4$,
because we can not calculate $k_{x,\mu}$ at
  positive sides of the boundary $\partial V$ due to the definition of (\ref{definition_of_k}) based on the forward lattice derivative.)

\begin{table}[htbp]
 \begin{center}
  \begin{tabular}{c||cccc||c}
   $\rho$&$|k_{x,\mu}|>1$&$k_{x,\mu}=-1$&$k_{x,\mu}=0$&$k_{x,\mu}=1$&$Q_V$\\
   \hline
    5&0&4&59105336&4&0.9674\\
   \hline
    10&0&8&59105328&8&0.9804\\
   \hline
    15&0&12&59105320&12&0.9490\\
   \hline
    20&0&12&59105320&12&0.8836
  \end{tabular}
 \end{center}
 \caption{The distribution of $k_{x,\mu}$ and the instanton charge $Q_V$ for a given $r$.}
 \label{result_one}
\end{table}

The results are summarized in TABLE \ref{result_one} and FIG. \ref{fig:one_loop}.
In our calculations of the monopole current configuration, we fix the center on the origin
\begin{equation}
 (b^1,b^2,b^3,b^4)=(0,0,0,0),
\end{equation}
and  change the value of $\rho$.

For a choice of $L=31$, the total number of configurations $\{ k_{x,\mu} \}$ are $62^4\times 4=59105344$.
Although the current $k_{x,\mu}$ is zero on almost all the links $(x,\mu)$, it has  a non-zero value $|k_{x,\mu}|=1$ on a small number of links, e.g., 4+4 links for $\rho=5a$.  
The number of links with $k_{x,\mu}=+1$ is equal to one with $k_{x,\mu}=-1$, which reflects the fact that the current $k_{x,\mu}$ draws a closed path of links.  
It turns out that there are no configurations such that $|k_{x,\mu}|>1$.

We see how $Q_V$ reproduces the total instanton charge $Q_P=1$ for the choice of $\rho$.
The instanton charge density is equal to zero when ${\bf A}_\mu(x)$ is a pure gauge, i.e., ${\bf F}_{\mu\nu}(x)=0$. 
Therefore, the more rapidly ${\bf A}_\mu(x)$ converge to a pure gauge, the more precisely $Q_V$ reproduces the proper value $Q_P$ for a given $L$.
For  one-instanton in the regular gauge, clearly, ${\bf A}_\mu(x)$ more rapidly converge to a pure gauge for smaller $\rho$.
Indeed, $Q_V$ for $\rho=10a$ reproduces $Q_P=1$ more precisely than  $\rho=15a$ and $\rho=20a$.
On the contrary, $Q_V$ for $\rho=5a$  reproduces $Q_P=1$  poorly compared with $\rho=10a$.  This is because the lattice is too coarse (a lattice spacing is not sufficiently small)  to estimate properly the rapid change of the instanton charge distribution concentrated to the neighborhood of the origin for small $\rho$.

If ${\bf A}_\mu(x)$ converge rapidly to a pure gauge $ih^\dagger(x)\partial_\mu h(x)$ so that $Q_V$ gives a good approximation for an expected value $Q_P$, 
then the color field  is expected to behave  as
${\bf n}(x) \rightarrow h^\dagger(x)T_3 h(x)$ asymptotically
and our choice of the boundary condition  
${\bf n}_x^\text{bound} :=h^\dagger(x)T_3 h(x)$ at the boundary $x \in \partial V$  is well motivated.

In FIG. \ref{fig:one_loop},  the support of $k_{x,\mu}$ is drawn by projecting the four-dimensional space on the $x_4=0$ hyperplane (3-dimensional space) for the choice of $\rho=5a, 10a, 15a$  and $20a$.
This figure shows that the non-zero monopole current forms a small loop.
The size of the magnetic monopole loop hardly change while $\rho$ increase.
This is an indication that the magnetic monopole loop  for one-instanton solution disappears in the continuum limit of the lattice spacing $a$ going to zero.

\section{Two instanton of the Jackiw-Nohl-Rebbi (JNR) type}

\begin{figure*}[htbp]
 \unitlength=0.001in
 \begin{picture}(7000,4600)(0,0)
  \put(0,2600){\includegraphics[trim=0 0 0 0, width=90mm]%
              {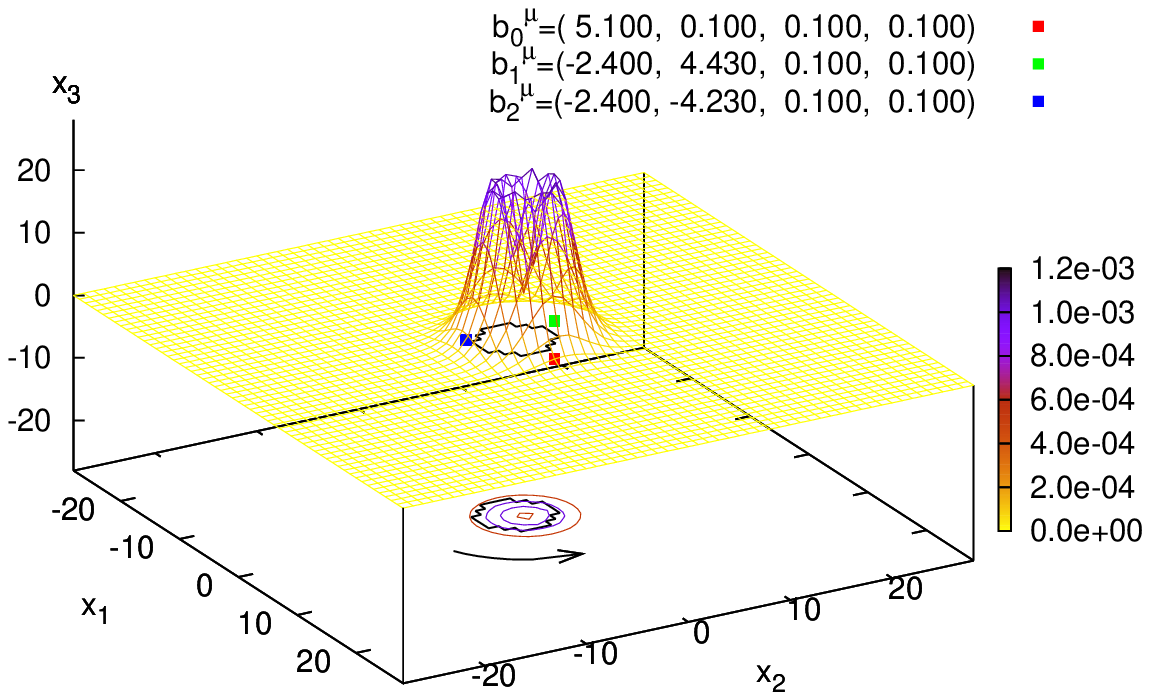}}%
  \put(1600,2600){(a)}
  \put(3500,2600){\includegraphics[trim=0 0 0 0, width=90mm]%
                 {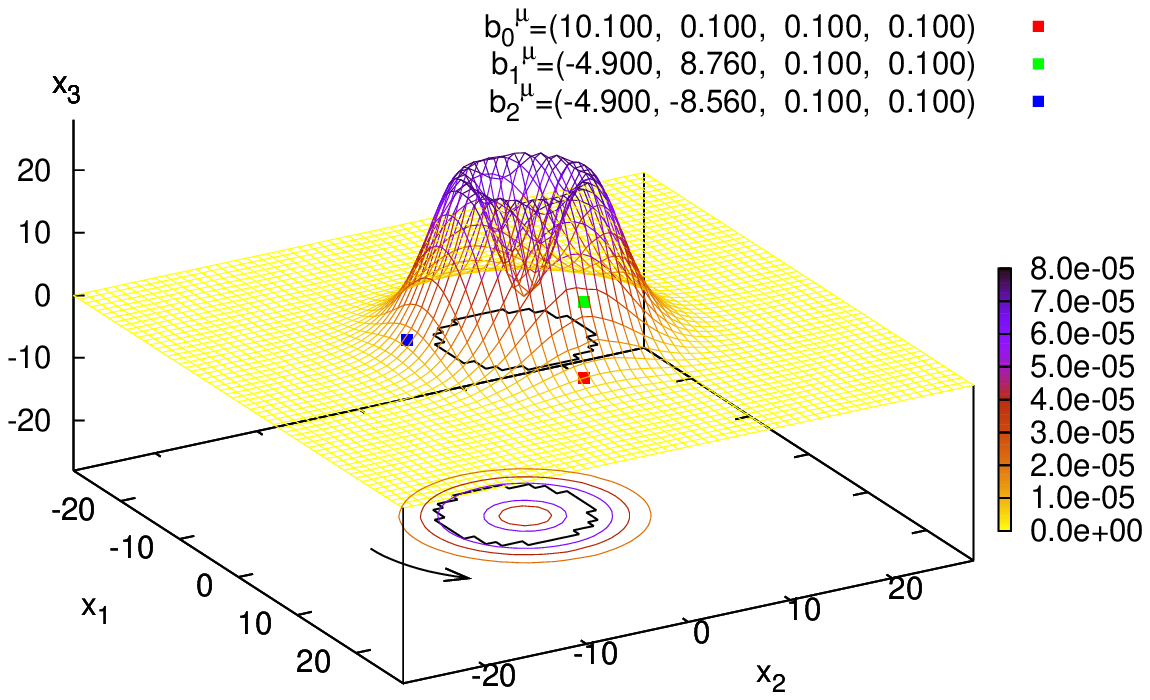}}%
  \put(5100,2600){(b)}
  \put(0,150){\includegraphics[trim=0 0 0 0, width=90mm]%
           {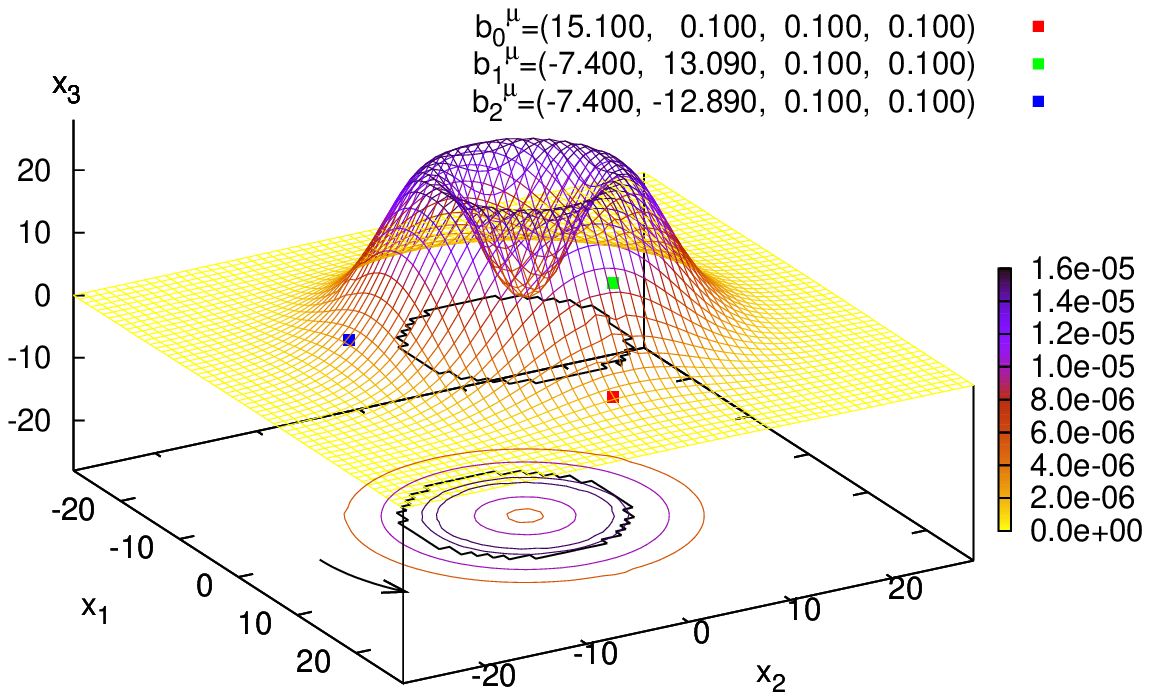}}%
  \put(1600,150){(c)}
  \put(3500,150){\includegraphics[trim=0 0 0 0, width=90mm]%
              {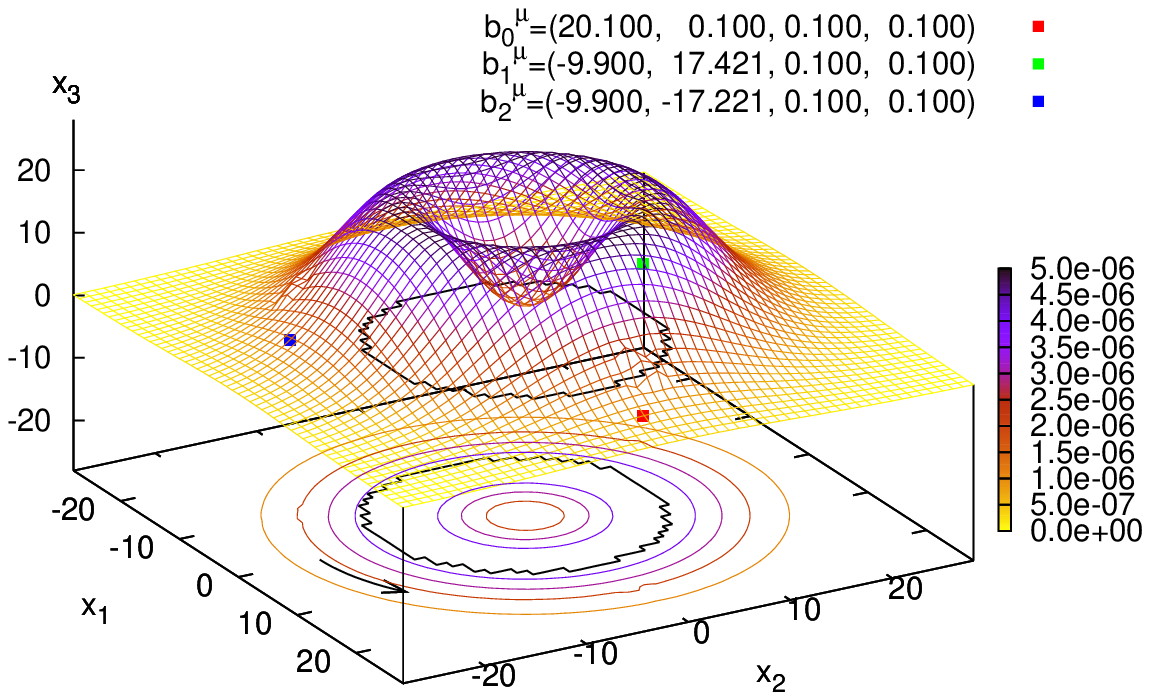}}%
  \put(5100,150){(d)}
 \end{picture}
 \caption{JNR two-instanton and the associated circular loop of the magnetic monopole current $k_{x,\mu}$.
The JNR two-instanton is defined by fixing three scales   $\rho_0=\rho_1=\rho_2=3a$ and three pole positions $b_0^\mu,b_1^\mu,b_2^\mu$  which are arranged to be three vertices of an equilateral triangle specified by $r$:   
 (a) $r=5a$, (b) $r=10a$, (c) $r=15a$ and (d) $r=20a$.
The grid shows an instanton charge density $D_x$ on $x_1$-$x_2$ ($x_3=x_4=0$) plane. 
The associated circular loop of the magnetic monopole current is located on the same plane as that specified by three poles. 
The black line on the base shows the magnetic monopole loop projected on the $x_1$-$x_2$ plane and the arrow indicates the direction of the monopole current, while colored lines on the base show the contour plot for the equi-$D_x$ lines.
Figures are drawn in units of $a$.
}
 \label{fig:JNR_loop}
\end{figure*}

\begin{figure*}[htbp]
 \unitlength=0.001in
 \begin{picture}(7000,3120)(0,0)
  \put(0,50){\includegraphics[trim=50 30 0 0, width=110mm]%
                              {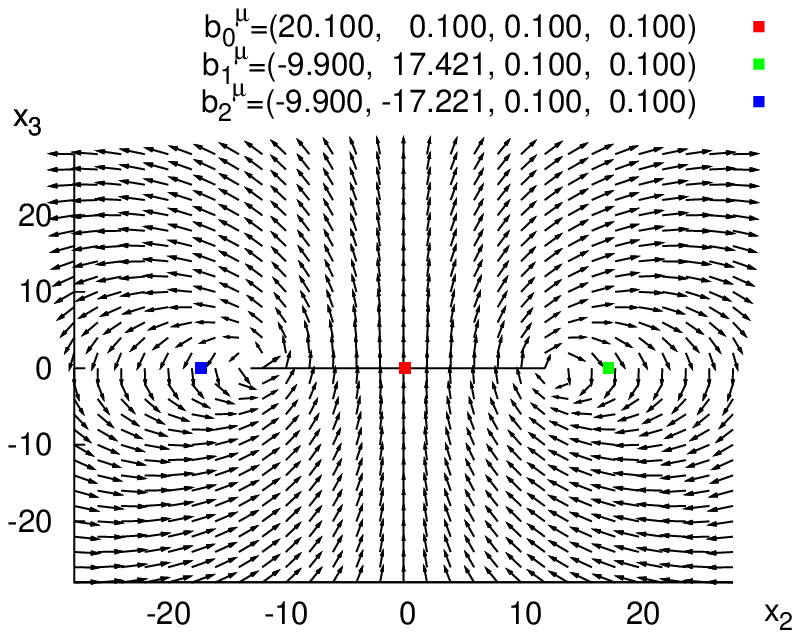}}
  \put(1600,10){(a)}
  \put(3500,50){\includegraphics[trim=50 30 0 0, width=110mm]%
                                {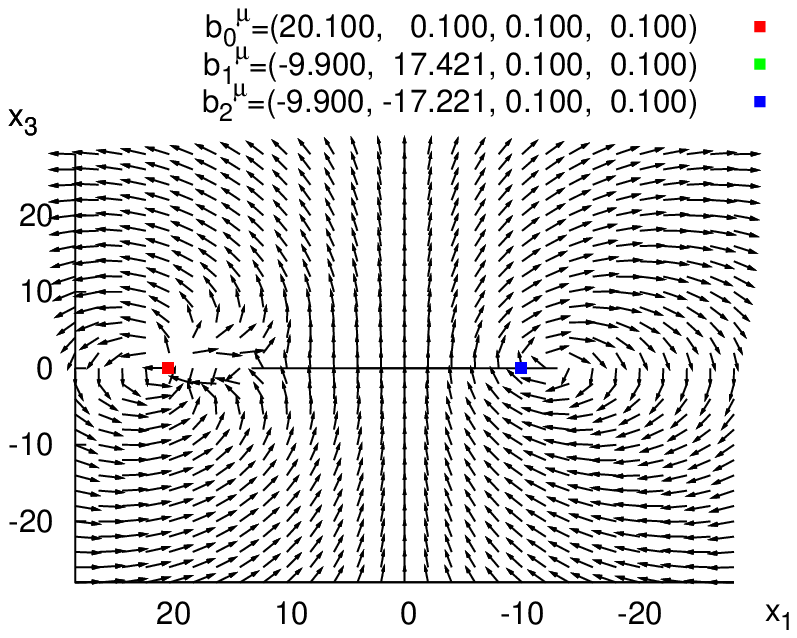}}
  \put(5100,10){(b)}
 \end{picture}
 \caption{
The configuration of the color field ${\bf n}_x=(n_x^1,n_x^2,n_x^3)$ and a  circular loop of the magnetic monopole current $k_{x,\mu}$  obtained from the JNR two-instanton solution ($r=20a$),  viewed in 
(a) the $x_2$-$x_3$ ($x_1=x_4=0$) plane which is off three poles, and 
(b) the $x_1$-$x_3$ ($x_2=x_4=0$) plane which goes through a pole $b_0^\mu$.
The magnetic monopole current $k_{x,\mu}$  and the three poles of the JNR solution are projected on the same plane.
Here the $SU(2)$ color field $(n_x^1,n_x^2,n_x^3)$ is identified with a unit vector in the three-dimensional space $(x_1,x_2,x_3)$. 
Figures are drawn in units of $a$.
}
 \label{fig:JNR_n}
\end{figure*}

It is known by the ADHM construction \cite{ADHM78} that the $N$-instanton moduli space has dimension $8N$.  
For $N=1$, the 8 moduli parameters are interpreted as 4+1+3 degrees of freedom for the position, size and SU(2) orientation (global gauge rotations), respectively.  

For $N=2$, the  Jackiw-Nohl-Rebbi (JNR) instanton \cite{JNR77} is the most general charge 2 instanton as explained below.   
The explicit form of the JNR two-instanton solution is 
\begin{align}
 g{\bf A}_\mu(x) =& - T^A \eta_{\mu\nu}^{A(-)} \partial_\nu \ln \phi_{\rm JNR}
 \\
=& T^A \eta_{\mu\nu}^{A(-)}
                  \phi_{\rm JNR}^{-1}
                  \sum_{r=0}^2\frac{2\rho_r^2\left(x^\nu-b^\nu_r\right)}
                                   {(|x-b_r|^2)^2} , 
\\
 \phi_{\rm JNR} :=& \sum_{r=0}^2\frac{\rho_r^2}{|x-b_r|^2} ,\ 
\end{align}
The JNR two-instanton has $4 \times 3 + 3 +3=18$ parameters, which consist of  three pole positions
$(b_0^1,b_0^2,b_0^3,b_0^4),(b_1^1,b_1^2,b_1^3,b_1^4),(b_2^1,b_2^2,b_2^3,b_2^4)$ 
and three scale parameters $\rho_0,\rho_1,\rho_2$ including the overall $SU(2)$ orientation.
Note that the number of poles is one greater than the number of the instanton charge. 
Although the parameter count of the JNR two-instanton appears to exceed the 16 dimensions of the $N=2$ moduli space,  
the JNR two-instanton has precisely the required number of parameters for the $N=2$ general solution. 
In fact, one parameter is reduced by noting that the multiplication of the scale parameter by a constant does not alter the solution, so only the ratios $\rho_r/\rho_0$ ($r=1,2$) are relevant.  
Moreover, one of the degrees of freedom corresponds to a gauge transformation \cite{JNR77}.

The 't Hooft two-instanton which is more popular and has been used extensively in the preceding investigations is given by 
\begin{equation}
  \phi_{\rm tHooft} := 1 + \sum_{r=1}^2\frac{\rho_r^2}{|x-b_r|^2} 
\end{equation}
The 't Hooft two-instanton has only $4 \times 2 + 2 +3=13$ parameters, which consist of  two pole positions
$(b_1^1,b_1^2,b_1^3,b_1^4),(b_2^1,b_2^2,b_2^3,b_2^4)$ 
and two scale parameters $\rho_1,\rho_2$ including the overall $SU(2)$ orientation.
The 't Hooft solution is reproduced from the JNR solution in the limit  $\rho_0=|b_0| \rightarrow \infty$, namely, the location of the first pole $b_0$ is sent to infinity keeping the relation $\rho_0=|b_0|$. 

The crucial difference between the JNR and the 't Hooft solutions is the asymptotic behavior.  
The JNR solution goes to zero slowly, while the gauge potential produced by the 't Hooft ansatz tends rapidly to zero at spatial infinity $|x| \rightarrow \infty$.  In fact, for $N=1$ the 't Hooft ansatz gives the one-instanton in the singular gauge with the asymptotic behavior 
\begin{equation}
{\bf A}_\mu(x) \sim O(|x|^{-3}) \quad |x| \rightarrow \infty ,
\end{equation}
while the one-instanton in the regular gauge exhibits the asymptotic behavior 
\begin{equation}
{\bf A}_\mu(x) \sim O(|x|^{-1}) \quad |x| \rightarrow \infty .
\end{equation}

In the case of the JNR two-instanton, we obtain
\begin{equation}
 h(x)=\frac{x_\mu}{|x|}\bar{e}_\mu,\ 
 \left(\bar{e}_\mu\equiv e_\mu^\dagger=(i\sigma_i,\bm{1})\right)
\end{equation}
and
\begin{align}
 &h^\dagger(x)T_3 h(x)=\frac{2\left(x_1x_3+x_2x_4\right)}{x^2}T_1\notag\\
 &\hspace{3cm}+\frac{2\left(-x_1x_4+x_2x_3\right)}{x^2}T_2\notag\\
 &\hspace{3cm}+\frac{-x_1^2-x_2^2+x_3^2+x_4^2}{x^2}T_3.
\end{align}
This is another form of the standard Hopf map \cite{Hopf31}.

We equate three size parameters
and put three pole positions
$(b_0^1,b_0^2,b_0^3,b_0^4)$, $(b_1^1,b_1^2,b_1^3,b_1^4)$,
$(b_2^1,b_2^2,b_2^3,b_2^4)$
on the $x_3=x_4=0$ plane,  
so that the three poles are located at the vertices of an  equilateral triangle:
\begin{align}
 \rho_0=& \rho_1=\rho_2 \equiv \rho ,
\\
 (b_0^1,b_0^2,b_0^3,b_0^4)=& (r,0,0,0)+\Delta ,
\\
 (b_1^1,b_1^2,b_1^3,b_1^4)
 =& \left(-\frac{r}{2},\frac{\sqrt{3}}{2}r,0,0\right)+\Delta ,
\\
 (b_2^1,b_2^2,b_2^3,b_2^4)
 =& \left(-\frac{r}{2},-\frac{\sqrt{3}}{2}r,0,0\right)+\Delta ,
\end{align}
where $\Delta$ is a small parameter introduced to avoid the pole singularities at $x=b_r$.

\begin{table}[htbp]
 \begin{center}
\begin{tabular}{c||cccc||c||c}
  $r$&$|k_{x,\mu}|>1$&$k_{x,\mu}=-1$&$k_{x,\mu}=0$&$k_{x,\mu}=1$&$l/r$&$Q_V$\\
  \hline
   5&0&14&59105316&14&5.6&1.903\\
  \hline
   10&0&26&59105292&26&5.2&1.969\\
  \hline
   15&0&40&59105264&40&5.3&1.950\\
  \hline
   20&0&51&59105242&51&5.1&1.862\\
\end{tabular} 
 \end{center}
 \caption{The distribution of $k_{x,\mu}$ and the instanton charge $Q_V$ for a given $r$.}
 \label{result_JNR}
\end{table}

In the numerical calculation, we choose
\begin{equation}
 \Delta=(0.1a,\ 0.1a,\ 0.1a,\ 0.1a) .
\end{equation}
Then we have searched for monopole currents by changing $r$. 
We have checked that for a given size of the lattice $L$ there is a suitable range of $\rho$, in which the effect of the discretization is sufficiently small.  
In view of this,  we adopt $L=31$ and $\rho=3a$. 

The results are summarized in TABLE \ref{result_JNR}
and FIG. \ref{fig:JNR_loop} and \ref{fig:JNR_n}.
In FIG. \ref{fig:JNR_loop}, as in the case of one-instanton in the regular gauge,
we draw the distribution of the instanton charge density $D_x$ and the support of the magnetic monopole current $k_{x,\mu}$ projected  on the $x_4=0$ hyperspace for $r=5a,10a,15a,20a$  
and $\rho=3a$.

The instanton charge density $D_x$ of the JNR two-instanton takes the maximal value at a circle with radius $R_I$, rather than on the origin,  on the $x_1$-$x_2$ plane.
This is not the case for the 't Hooft two instanton in which the instanton charge distribution concentrates near the two pole positions, as is well known.

We have found that non-vanishing monopole currents originating from the JNR two-instanton forms a  circular loop.
The circular loops of the magnetic monopole current are located on the same plane as that specified by three poles $b_0,b_1,b_2$.
The size of the circular loop, e.g., the radius $R$, increases proportionally as $r$ increases and the circular loops constitute concentric circles with the center at the origin, within the accuracy of our numerical calculations. 

To reproduce the correct distribution of the instanton charge density $D_x$ in the numerical calculations, we need to  further approximate the link variable $U_{x,\mu}$ defined by (\ref{U}) using a discretization with a width $\epsilon=a/n$ for the line integral on a link:
\begin{align}
 U_{x,\mu}&\thickapprox
            \exp\Bigg\{i \epsilon\sum_{k=1}^{n}
                       \frac{1}{2}
                       \left[g{\bf A}_\mu(x+(k-1)\epsilon\hat{\mu})\right.\notag\\
          &\left.\hspace{3cm}+g{\bf A}_\mu(x+k\epsilon\hat{\mu})
                       \right]
                \Bigg\} .
 \label{def_of_U}
\end{align}
We have chosen the lattice spacing $a=1$ and the total number of partition points $n=20$.
If $n$ is relatively small, the numerical result for the instanton charge density $D_x$ gives a wrong distribution with remnant of three poles.

In TABLE \ref{result_JNR}, we find that the ratio between the length $\ell := \sum_{x,\mu} |k_{x,\mu}|$ of the magnetic monopole current $k_{x,\mu}$ and the size  $r$ of the equilateral triangle of JNR is nearly constant, 
\begin{equation}
 \ell/r \simeq 5.2  ,
\end{equation}
which  implies  
\begin{equation}
  R_m/r \simeq 0.65  ,
\end{equation}
where we have used a relation $\ell \simeq 8 R_m$ for large $\ell$ since a closed current $k_{x,\mu}$ consists of links on a lattice and the relation $\ell \simeq 2\pi R_m$ in the continuum does not hold. 
Moreover, the magnetic monopole loop passes along the neighborhood of contour giving the absolute maxima of the instanton charge density,  
\begin{equation}
 R_m/R_I \simeq 0.65/0.54 \simeq  1.2 , 
\end{equation}
where $R_I \simeq 0.54r$.

FIG. \ref{fig:JNR_n} shows the relationship between the magnetic monopole loop $k_{x,\mu}$ and the color field ${\bf n}_x$ configuration.
The vector field $\{{\bf n}_x\}$ is winding around the loop,
and it is indeterminate at points where the loop pass.
The configurations of the color field giving the magnetic monopole loop were made available for the first time in this study based on the new reformulation of Yang-Mills theory.

\section{Conclusion and discussion}\label{sec:conclusion}

For given instanton solutions of the classical  Yang-Mills equation in the four-dimensional Euclidean space, we have solved in a numerical way the reduction condition to obtain the color field which plays the key role to define a gauge-invariant magnetic monopole in our reformulation of the Yang-Mills theory written in terms of new variables.  
Here we have used a lattice regularization \cite{KKMSSI05} for performing numerical calculations.
Then we have constructed the magnetic monopole current $k_\mu$ on a dual lattice where the resulting magnetic charge is gauge invariant and quantized according to the quantization condition of the Dirac type. 

For the two-instanton solution of the Jackiw-Nohl-Rebbi type, we have discovered that the magnetic monopole current $k_\mu$ has the support on a circular loop which is located near the maxima of the instanton charge density.  
Thus, we have shown that the two-instanton solution of the Jackiw-Nohl-Rebbi type generates the magnetic monopole loop in four-dimensional SU(2) Yang-Mills theory. 
In the same setting, we have found that the magnetic current has the support only on a plaquette around the center of the one-instanton.  This result confirms that no magnetic monopole loop is generated for one-instanton solution in the continuum limit. 
 
Combining the result in this paper with the previous one \cite{KFSS08},  we have found that both the JNR two-instanton solution and two-merons solution with the same asymptotic behavior at spacial infinity: 
\begin{equation}
{\bf A}_\mu(x) \sim O(|x|^{-1}) \quad |x| \rightarrow \infty 
\end{equation}
generate  circular loops of magnetic monopole, which should be compared with the 't Hooft (multi) instanton with the asymptotic behavior at spacial infinity: 
\begin{equation}
{\bf A}_\mu(x) \sim O(|x|^{-3}) \quad |x| \rightarrow \infty .
\end{equation}
We expect that these loops of magnetic monopole are responsible for confinement in the dual superconductivity picture. 
This result seems to be consistent with the claim made in \cite{LNT04,Lenz09}.  
However, the correspondence between the instanton and magnetic monopole loop is not one-to-one.  
To draw the final conclusion, we need to collect more data for supporting this claim.
Moreover, it is not yet clear which relationship between instanton charge and the magnetic charge holds in our case, as studied in \cite{Reinhardt97,Jahn00,TTF00}.
These issues will be investigated in future works.

\section*{Acknowledgements} 
The authors would like to thank Koji Hashimoto for very helpful discussions on instantons in the early stage of this work. 
This work is financially supported by Grant-in-Aid for Scientific Research (C) 21540256 from Japan Society for the Promotion of Science
(JSPS).

\appendix
\section{Deriving the reduction equation on a lattice}
Taking the square root of both sides of \eqref{RDEonLattice1} yields
\begin{equation}
 \lambda_x=\pm\sqrt{W^A_xW^A_x} ,
\end{equation}
which is substituted into \eqref{RDEonLattice1} to obtain  
\begin{equation}
 n_x^A=\pm\frac{W^A_x}{\sqrt{W^B_xW^B_x}} .
\label{RDEonLattice3}
\end{equation}
We must choose a correct sign in the right-hand side of the equation \eqref{RDEonLattice3} so that a solution of this equation gives a solution of the original equation \eqref{RDEonLattice1}.

In what follows, we adopt a direct method of choosing a correct sign of the right-hand side by examining whether the equation \eqref{RDEonLattice3} converges to the RDE in the continuum limit $a \rightarrow 0$.
By substituting $U_{x,\mu}=e^{-iag{\bf A}_\mu(x)}$ into \eqref{W_x^A} and expanding it in powers of $a$, $W_x^A$ is written as
\begin{widetext}
\begin{align}
 W_x^A
 &=4\sum_{\mu=1}^4\text{tr}
    \left[\left\{e^{-iag{\bf A}_\mu(x)}
                 {\bf n}(x+a\hat{\mu})
                 e^{iag{\bf A}_\mu(x)}
                 +e^{iag{\bf A}_\mu(x-a\hat{\mu})}
                  {\bf n}(x-a\hat{\mu})
                  e^{-iag{\bf A}_\mu(x-a\hat{\mu})}
          \right\}T^A
    \right]/\text{tr}\left({\bf 1}\right)\notag\\
 &=4\sum_{\mu=1}^4\text{tr}
    \left[\left\{{\bf n}(x+a\hat{\mu})
                 -ia\left[g{\bf A}_\mu(x),
                          {\bf n}(x+a\hat{\mu})
                    \right]\vphantom{\frac{a^2}{2}}
                 -\frac{a^2}{2}
                  \left[g{\bf A}_\mu(x),
                        \left[g{\bf A}_\mu(x),{\bf n}(x+a\hat{\mu})\right]
                  \right]\right.\right.\notag\\
 &\hspace{3cm} +{\bf n}(x-a\hat{\mu})
                 +ia\left[g{\bf A}_\mu(x-a\hat{\mu}),{\bf n}(x-a\hat{\mu})
                    \right]\notag\\[+1mm]
 &\hspace{5cm}\left.\left.
                 -\frac{a^2}{2}
                  \left[g{\bf A}_\mu(x-a\hat{\mu}),
                        \left[g{\bf A}_\mu(x-a\hat{\mu}),
                              {\bf n}(x-a\hat{\mu})
                        \right]
                  \right]
          \right\}T^A
    \right]/\text{tr}\left({\bf 1}\right)+O(a^3)\notag\\
 &=4\sum_{\mu=1}^4\text{tr}
    \left[\left\{2{\bf n}(x)
                 +a^2\partial_\mu\partial_\mu{\bf n}(x)
          \right.\right.\notag\\[-3mm]
 &\hspace{2.5cm}   -2ia^2\left[g{\bf A}_\mu(x),\partial_\mu{\bf n}(x)\right]
                 -ia^2\left[g\partial_\mu{\bf A}_\mu(x),
                            {\bf n}(x)
                      \right]\notag\\[+2mm]
 &\hspace{5.8cm}\left.\left.
                 -a^2
                  \left[g{\bf A}_\mu(x),
                        \left[g{\bf A}_\mu(x),{\bf n}(x)\right]
                  \right]
          \right\}T^A
    \right]/\text{tr}\left({\bf 1}\right)+O(a^3)\notag\\
 &=4\sum_{\mu=1}^4\text{tr}
    \left[\left\{2{\bf n}(x)
                 +a^2D_\mu[{\bf A}]D_\mu[{\bf A}]{\bf n}(x)
          \right\}T^A
    \right]/\text{tr}\left({\bf 1}\right)+O(a^3)\notag\\
 &=8n^A(x)
   +a^2\sum_{\mu=1}^4
       \left(D_\mu[{\bf A}]D_\mu[{\bf A}]{\bf n}(x)\right)^A
   +O(a^3).
   \label{W}
\end{align}
Then, the magnitude of $W_x^A$ is given by 
\begin{align}
 \sqrt{W^A_xW^A_x}
 &=\sqrt{64+16a^2\sum_{\mu}n^A(x)
               \left(D_\mu[{\bf A}]D_\mu[{\bf A}]{\bf n}(x)
               \right)^A
          +O(a^3)}\notag\\
 &=8+a^2\sum_{\mu}
     n^A(x)\left(D_\mu[{\bf A}]D_\mu[{\bf A}]{\bf n}(x)\right)^A
    +O(a^3).
    \label{sqrtW}
\end{align}
Substituting \eqref{W} and \eqref{sqrtW} into \eqref{RDEonLattice3}, therefore, we obtain
\begin{align}
 8\left(-n^A(x)\pm n^A(x)\right)
 +a^2\left(-\sum_{\mu=1}^4
  \left(D_\mu[{\bf A}]D_\mu[{\bf A}]{\bf n}(x)\right)^A
        \pm n^A(x)\sum_{\mu}n^B(x)
            \left(D_\mu[{\bf A}]D_\mu[{\bf A}]{\bf n}(x)\right)^B
  \right)
 +O(a^3)=0 ,
\label{EXPANSIONofRDEonLattice}
\end{align}
\end{widetext}
where the double-signs $\pm$ in \eqref{RDEonLattice3} and \eqref{EXPANSIONofRDEonLattice}
correspond to each other. 
If we choose the minus sign, we have an unrealistic result $n^A(x)=0$. 
Thus, by choosing the plus sign in \eqref{RDEonLattice2}, the leading term vanishes and the next-to-leading term leads to 
\begin{gather}
 \sum_{\mu=1}^4\left(D_\mu[{\bf A}]D_\mu[{\bf A}]{\bf n}(x)\right)^A
 =-\lambda(x)n^A(x)\\
 \lambda(x)\equiv -\sum_{\mu}n^B(x)
                 \left(D_\mu[{\bf A}]D_\mu[{\bf A}]
                       {\bf n}(x)
                 \right)^B ,
\end{gather}
which is nothing but the RDE in the continuum theory:
\begin{equation}
 -D_\mu[{\bf A}]D_\mu[{\bf A}]{\bf n}(x)=\lambda(x){\bf n}(x) .
\end{equation}



\begin{thebibliography}{99}
\bibitem{Wilson74}
  K. Wilson, 
  Phys. Rev. D {\bf 10}, 2445--2459
 (1974).


\bibitem{YM54}
  C.N. Yang and R.L. Mills,
Phys. Rev. {\bf 96}, 191--195
 (1954).


\bibitem{MS}
N. Manton and P. Sutcliffe,
Topological Solitons
(Cambridge Univ. Press, 2007).
\\
R. Rajaraman,
Solitons and Instantons: An Introduction to Solitons and Instantons in Quantum Field Theory
(North-Holland, Amsterdam, 1987).


\bibitem{BPST75}
A.A. Belavin, A. M. Polyakov, A.S. Shwarts and  Yu.S. Tyupkin,
Phys.Lett. B {\bf 59}, 85--87 (1975). 


\bibitem{CF77}
E. Corrigan and D.B. Fairlie,
Phys. Lett. B {\bf 67}, 69 (1977).


\bibitem{tHooft76}
G. 't Hooft,
Phys.Rev. D{\bf 14}, 3432-3450 (1976); Erratum-ibid. D{\bf 18}, 2199 (1978). 


\bibitem{Wilczek77}
F. Wilczek,
in ``Quark Confinement and Field Theory'' eds. by D. Stump and D. Weingarten, (Wiley, New York, 1977).


\bibitem{Witten79}
E. Witten,
Nucl. Phys. B {\bf 149}, 285--320 (1979). 


\bibitem{JNR77}
R. Jackiw, C. Nohl and C. Rebbi,
Phys.Rev. D{\bf 15}, 1642--1646 (1977).

\bibitem{ADHM78}
M.F. Atiyah, N.J. Hitchin, V.G. Drinfeld and Y.I. Manin, 
Phys. Lett. A {\bf 65}, 185 (1978).


\bibitem{AFF76}
V. De Alfaro, S. Fubini and G. Furlan,
Phys. Lett. B {\bf 65}, 163--166
 (1976).


\bibitem{AFF77}
V. De Alfaro, S. Fubini and G. Furlan,
Phys. Lett. B {\bf 72}, 203--207
 (1977).


\bibitem{Actor79}
A. Actor,
Rev. Mod. Phys. {\bf 51}, 461--525 (1979).


\bibitem{CDG78}
 C.G. Callan, Jr., R. Dashen and D.J. Gross,
Phys. Rev. D {\bf 17}, 2717--2763 (1978).
\\
C.G. Callan, Jr., R.F. Dashen and  D.J. Gross, 
Phys.Lett. B{\bf 66}, 375--381 (1977). 

\bibitem{dualsuper}
  Y. Nambu,
  Phys. Rev. D {\bf 10}, 4262--4268
 (1974).
\\
G. 't Hooft,
  in: High Energy Physics, edited by A. Zichichi 
(Editorice Compositori, Bologna, 1975).
\\
S. Mandelstam,
 Phys. Report  {\bf 23}, 245--249
 (1976).


\bibitem{Polyakov77}
A.M. Polyakov,
  Phys. Lett. B {\bf 59}, 82--84
 (1975).
 \\
A.M. Polyakov,
  Nucl. Phys. B {\bf 120}, 429--458
 (1977).


\bibitem{Kondo08b}
K.-I. Kondo,  
arXiv:0802.3829 [hep-th], 
J.Phys. G{\bf 35}, 085001 (2008). 


\bibitem{KFSS08}
K.-I.  Kondo, N. Fukui, A. Shibata and T. Shinohara,
arXiv:0806.3913 [hep-th],
Phys.Rev.D{\bf 78}, 065033 (2008). 
\\
K.-I.  Kondo,
e-Print: arXiv:0812.4026 [hep-th], 
PoS CONFINEMENT8, 046 (2008). 


\bibitem{SKKISF09}
A. Shibata, K.-I. Kondo, S. Kato, S. Ito,  T. Shinohara and N. Fukui,  
Talk given at 27th International Symposium on Lattice Field Theory (Lattice 2009), Beijing, China, 25-31 Jul 2009. 
arXiv:0911.4533 [hep-lat] 


\bibitem{CG95}
M.N. Chernodub and F.V. Gubarev,
[hep-th/9506026],
JETP Lett. {\bf 62}, 100 (1995).


\bibitem{BOT97}
R.C. Brower, K.N. Orginos and C-I. Tan,
[hep-th/9610101], 
Phys. Rev. D {\bf 55}, 6313--6326 (1997).
\\
R.C. Brower, K.N. Orginos and C-I. Tan,
hep-lat/9608012.


\bibitem{BHVW01}
F. Bruckmann, T. Heinzl, T. Vekua and A. Wipf,
[hep-th/0007119],
Nucl. Phys. B {\bf 593}, 545--561 (2001).
\\
F. Bruckmann,
[hep-th/0011249],
JHEP 08, 030 (2001).



\bibitem{MN02}
A. Montero and J.W. Negele,
[hep-lat/0202023],
Phys. Lett. B {\bf 533}, 322
 (2002).
\\
J.V. Steele and J.W. Negele,
[hep-lat/0007006],
Phys. Rev. Lett. {\bf 85}, 4207--4210 (2000).


\bibitem{HT96}
A. Hart and M. Teper,
[hep-lat/9511016],
Phys. Lett. B {\bf 371}, 261--269 (1996).


\bibitem{STSM96}
H. Suganuma, A. Tanaka, S. Sasaki and O. Miyamura,  
e-Print: hep-lat/9512024,  
Nucl. Phys. Proc. Suppl. {\bf 47}, 302-305 (1996). 


\bibitem{RT01}
H. Reinhardt and T. Tok,
Phys. Lett. B {\bf 505}, 131--140 (2001).
\\
H. Reinhardt and T. Tok,
hep-th/0009205.


\bibitem{BH03}
F. Bruckmann and D. Hansen,
[hep-th/0305012],
Ann. Phys. {\bf 308}, 201--210 (2003).


\bibitem{KL03}
N. D. Lambert and D. Tong, 
[hep-th/9907014],
Phys. Lett. B{\bf 462},  89-94 (1999). 
\\
S. Kim and K. Lee, 
[hep-th/0307048],
JHEP 0309, 035 (2003). 
\\
H.-Y. Chen,  M. Eto and K. Hashimoto, 
[hep-th/0609142], 
JHEP 0701, 017 (2007). 


\bibitem{tHP74}
G. 't Hooft,
Nucl. Phys. B{\bf 79}, 276-- (1974).
\\
A.M. Polyakov,
JETP Lett.{\bf 20}, 194-195 (1974), 
Pisma Zh.Eksp.Teor.Fiz. {\bf 20}, 430-433 (1974). 


\bibitem{tHooft81}
  G. 't Hooft,
  Nucl.Phys. B {\bf 190} [FS3], 455--478
 (1981).


\bibitem{KLSW87}
  A. Kronfeld, M. Laursen, G. Schierholz and U.-J. Wiese,
  Phys.Lett. B {\bf 198}, 516--520
 (1987).  


\bibitem{SY90} 
  T. Suzuki and I. Yotsuyanagi,
  Phys. Rev. D {\bf 42}, 4257--4260 (1990).


\bibitem{SNW94}
J.D. Stack, S.D. Neiman and R. Wensley,
[hep-lat/9404014],
Phys. Rev. D{\bf 50}, 3399--3405 (1994).
H.~Shiba and T.~Suzuki,
Phys. Lett. B{\bf 333}, 461--466  (1994).
  
  
\bibitem{AS99}
  K. Amemiya and H. Suganuma,
[hep-lat/9811035],
Phys. Rev. D{\bf 60}, 114509 (1999).
\\
H. Suganuma, K. Amemiya, H. Ichie, N. Ishii, H. Matsufuru and T.T. Takahashi,
[hep-lat/0407016], 
Nucl. Phys. B (Proc. Suppl.) {\bf 106}, 679--681 (2002)
\\
  V.G. Bornyakov, M.N. Chernodub, F.V. Gubarev, S.M. Morozov and M.I. Polikarpov, 
[hep-lat/0302002],
 Phys. Lett. B{\bf 559}, 214--222 (2003).
 

\bibitem{CP97}
M.N. Chernodub, M.I. Polikarpov,
hep-th/9710205.
\\
 J. Greensite,
[hep-lat/0301023],  
Prog. Part. Nucl. Phys. {\bf 51}, 1 (2003).  


\bibitem{Cho80}
  Y.M. Cho,
Phys. Rev. D {\bf 21}, 1080--1088 (1980).
Y.M. Cho,
Phys. Rev. D {\bf 23}, 2415--2426 (1981). 


\bibitem{DG79}
  Y.S. Duan and M.L. Ge, 
Sinica Sci., {\bf 11}, 1072--1081 (1979). 


\bibitem{FN99} 
  L. Faddeev and A.J. Niemi,
[hep-th/9807069],
Phys. Rev. Lett. {\bf 82}, 1624--1627 (1999).


\bibitem{Shabanov99}
  S.V. Shabanov,
[hep-th/9903223],
Phys. Lett. B {\bf 458}, 322--330 (1999).
\\
  S.V. Shabanov, 
[hep-th/9907182],
Phys. Lett. B {\bf 463}, 263--272 (1999).


\bibitem{KMS06}
  K.-I. Kondo, T. Murakami and T. Shinohara,
[hep-th/0504107], 
Prog. Theor. Phys. {\bf 115}, 201--216 (2006). 


\bibitem{KMS05}
  K.-I. Kondo, T. Murakami and T. Shinohara,
[hep-th/0504198],
Eur. Phys. J. C {\bf 42}, 475--481 (2005).


\bibitem{Kondo06}
K.-I. Kondo,
[hep-th/0609166], 
Phys. Rev. D {\bf 74}, 125003 (2006). 

 
\bibitem{KSM08}
K.-I. Kondo, T. Shinohara and T. Murakami,
e-Print: arXiv:0803.0176 [hep-th],
Prog. Theor. Phys. {\bf 120},  1--50 (2008).


\bibitem{KKMSSI05}
  S. Kato, K.-I. Kondo, T. Murakami, A. Shibata, T. Shinohara and S. Ito,
[hep-lat/0509069], 
Phys. Lett. B {\bf 632}, 326--332
 (2006).


\bibitem{IKKMSS06}
  S. Ito, S. Kato, K.-I. Kondo, T. Murakami, A. Shibata and T. Shinohara,  
[hep-lat/0604016], 
Phys. Lett. B {\bf 645}, 67--74  (2007).  
 

\bibitem{SKKMSI07}
A. Shibata, S. Kato, K.-I. Kondo, T. Murakami, T. Shinohara and  S. Ito,
arXiv:0706.2529 [hep-lat],
Phys.Lett. B{\bf 653}, 101--108 (2007). 


\bibitem{SKKMSI07b}
A. Shibata, S. Kato, K.-I. Kondo, T. Murakami, T. Shinohara, and S. Ito,
e-Print: arXiv:0710.3221 [hep-lat], 
POS(LATTICE-2007) 331.
\\
A. Shibata,  K.-I. Kondo, S. Kato, S. Ito, T.Shinohara, T. Murakami, 
arXiv:0810.0956 [hep-lat],
PoS(LATTICE 2008)268. 


\bibitem{KKSSI09}
S. Kato,  K.-I. Kondo, A. Shibata,  T. Shinohara, S. Ito, 
To appear in the proceedings of 27th International Symposium on Lattice Field Theory (Lattice 2009), Beijing, China, 25-31 Jul 2009,  
arXiv:0911.0755 [hep-lat]. 


\bibitem{KSSMKI08}
K.-I. Kondo, A. Shibata, T. Shinohara, T. Murakami, S. Kato and  S. Ito,
arXiv:0803.2451[hep-lat],
PLB{\bf 669}, 107--118 (2008).


\bibitem{SKS09}
A. Shibata, K.-I. Kondo and T. Shinohara,
arXiv:0911.5294 [hep-lat].


\bibitem{Cho00}
Y.M. Cho,
Phys. Rev. D {\bf 62}, 074009 (2000).


\bibitem{Kondo08}
  K.-I. Kondo,
arXiv:0801.1274 [hep-th],
Phys. Rev. D {\bf 77}, 085029 (2008).


\bibitem{KS08}
 K.-I. Kondo and A. Shibata, 
arXiv:0801.4203 [hep-th].


\bibitem{Hopf31}
H. Hopf,
%
Math. Ann. {\bf104}, 637 (1931).
\\
M. Minami,
Prog. Theor. Phys. {\bf 62}, 1128--1142 (1979).
\\
L.H. Ryder,
J. Phys. A{\bf 13}, 437 (1980).


\bibitem{LNT04}
F. Lenz, J.W. Negele and M. Thies,
[hep-th/0306105],
Phys.Rev.D {\bf 69}, 074009 (2004).


\bibitem{Lenz09}
F. Lenz,
e-Print: arXiv:0909.3290 [hep-ph],
Int. J. Mod. Phys. A{\bf 25}, 490--501 (2010). 


\bibitem{Reinhardt97}
H. Reinhardt,
[hep-th/9702049],
Nucl. Phys. B {\bf 503}, 505--529 (1997).


\bibitem{Jahn00}
O. Jahn,
[hep-th/9909004],
J. Phys. A {\bf 33}, 2997--3019 (2000).


\bibitem{TTF00}
T. Tsurumaru, I. Tsutsui and A. Fujii,
[hep-th/0005064],





\end{thebibliography}
\end{document}